# *A Picture for the Words!*
# Textual Visualization in Big Data Analytics

***Cherilyn Conner*** *connerc3@student.wpunj.edu*
***Jim Samuel*** *jim@aiknowledgecenter.com*
***Andrey Kretinin*** *kertinina@wpunj.edu*
***Yana Samuel*** *Yana.Samuel@gmail.com*
***Lee Nadeau*** *Nadeaul@student.wpunj.edu*

**ABSTRACT**
Data Visualization has become an important aspect of big data analytics and has grown in sophistication and variety. We specifically identify the need for an analytical framework for data visualization with textual information. Data visualization is a powerful mechanism to represent data, but the usage of specific graphical representations needs to be better understood and classified to validate appropriate representation in the contexts of textual data and avoid distorted depictions of underlying textual data. We identify prominent textual data visualization approaches and discuss their characteristics. We discuss the use of multiple graph types in textual data visualization, including the use of quantity, sense, trend and context textual data visualization. We create an explanatory classification framework to position textual data visualization in a unique way so as to provide insights and assist in appropriate method or graphical representation classification.

## 1 INTRODUCTION
There has been significant increase in the gathering, storing, manipulation, representation and analysis of textual information in big data initiatives. Social media in particular has fueled the growth of textual analytics (Stieglitz, et., al., 2018). A significant increase in access to and use of mobile devices has also boosted social media activity (Jimenez-Marquez, Gonzalez- Carrasco, Lopez-Cuadrado, & Ruiz-Mezuca, 2019). Textual data is a rich source of information and lends itself to various forms of creative analysis, providing an opportunity for the application of big data techniques. Organizations are placing great emphasis on the gathering and analysis of textual information in looking for ways to improve their understanding of user sentiment, behavior, trust, loyalty and various other critical decision-influencing variables. Textual data has the potential to represent unique user information in the form of direct statements and latent implications which cannot be gathered from numerical data and other forms of data, and therefore has unique potential.
One of the main strategies used for textual data analysis and insight generation is the visualization of textual data. Data visualization in general has grown in popularity and the importance of data visualization has increased with the growth in big data. Data visualization is a quick and easy way to convey information and can be used for effective storytelling, enabling parsimonious ‘connection of dots’, identifying trends, finding outliers, generating audience interest and providing summary representations of various forms of data. Common examples of data visualization graphs and artifacts include various types of pie charts, histograms, wordclouds, bubble clouds, bar charts, tree maps, bubble charts, step charts, area charts, heat maps, cartograms, dot distribution maps, proportional symbol maps, timelines, time series, scatter plots, Gantt charts, stream graphs, polar areas, spider charts, box and whisker plots, tree visualizations, dendrograms, radial trees, spark lines, node-link diagrams, alluvial diagrams and matrices, amongst others. However, fewer plots and methods are available for textual analytics and even then, some of the interpretations are more difficult. Creative plot types of textual data using innovative analytics, such as wordclouds (Figure 1.) are useful representations, but additional classification is necessary for better understanding the contribution value of such visualizations.
In spite of the rapid advancement of data analytics and visualization technologies, there remain a few challenges for data visualization, especially in the context of big data. The classical challenge of information loss, the inadequacy of a summary representation to reflect important particulars, in the form of granular details remains (Matzen, et., al., 2018). Additionally, data visualization for big data requires extensive processing power and additional ontological challenges remain with the visualization of data from newly created variables. It must also be pointed out that not all visualization can be categorized as ‘data visualization’ – for example, visualization as infographics may narrate a premediated story or infographics may contain abstract concepts, time and process representations which are not directly data driven. Such infographics may contain data visualizations but are of themselves of a different genre of graphics. Data visualizations, on the other hand, are directly driven by data and serve as statistical or mathematical method-driven graphical representations of data, and thus are relatively more objective and reflective of the underlying data.
Traditionally the domain of data visualization has been dominated by numerical data, and this has simply been an extension of the fact that most traditional data analysis and statistics were focused on numerical data, with additional parameters to ensure quality of data. With the advent of big data, there has been a natural transition to various additional data types and a borrowing and adapting of data visualization methods. One of the most voluminous data types in big data has been textual data or “character” data. We are in the middle of a huge demand wave for textual analytics, with a particular focus on social media analytics and this creates a need for understanding textual data visualization better. Our study focuses on creating a framework to better understand and classify data visualization for textual analytics.

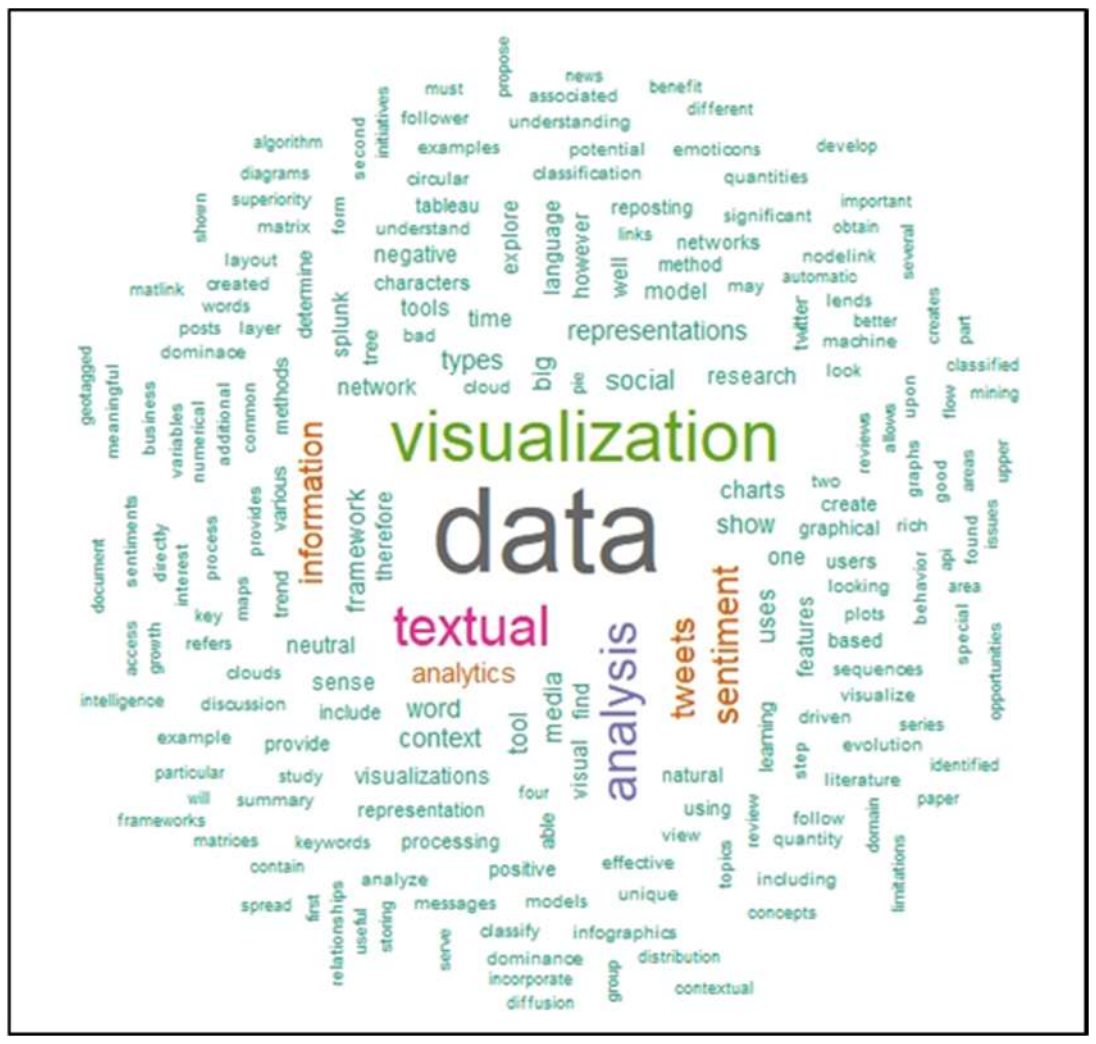

Figure 1: Wordcloud for this research paper

The sections that follow consist of a literature review covering quantity, sense, context and trend as the core focal points, framework presentation, discussion and conclusion.

## 2 LITERATURE REVIEW

The present literature review provides a summary of our review of past research that addresses textual data visualization. Social media data analysis and visualization provides rich opportunities to examine the use of visualization analytics. Amazon product reviews have been analyzed to find differences in reviews to explain customer behavior (Jimenez-Marquez, Gonzalez- Carrasco, Lopez-Cuadrado, & Ruiz-Mezuca, 2019). Hotel user reviews have also been used to capture the atmosphere and perceptions of the hotel. Textual analytics and visualization have been used to demonstrate links between news and financial markets, and to show evolution in politics including key players, issues, and topics (Chen, Lin, & Yuan, 2017). Textual analytics and visualization have also been used to help understand game stages or special events and visualize sub-event peaks, sentiment and keyword analysis for box-office forecasting, event distribution for urban planning and social movement of inter-city entities (Chen, Lin, & Yuan, 2017). A broader discussion of textual analytics would include audio data as well; however, we limit our focus to typed textual data. Our literature review and research on textual representations led us to classify four key quadrants for textual data visualization: quantity, sense, context and trend

### Quantities Visualization

Quantities Visualization (QViz) refers to the types of visual graphical representations of analysis that incorporate counts of textual characters, features, or sequences. Common examples include wordclouds, count frequencies, tables, and pie charts. Wordclouds can be used as an initial step in analysis before moving on to other methods (Kabir, Karim, Newaz & Hossain, 2018). A wordcloud displays a pictorial representation of frequently use words where word size is determined by the frequency of a word’s usage, therefore the more often a word is used the bigger it will be in the cloud. They are useful to discover the most frequently discussed topics and can be applied in a wide range of contexts. Kabir, Karim, Newaz, and Hossain (2018) were able to use a wordcloud to display the topics that a Bangladeshi politician tweeted about most. However, the use of a wordcloud takes away the context of the topic, therefore there is no indication if the topic is being talked about positively or negatively (Kabir, Karim, Newaz & Hossain, 2018). SentenTree is a visualization tool that can be used to help users understand concepts and opinions by visualizing frequent keyword patterns (Chen, Lin, & Yuan, 2017).
A notable example of the visualization of quantities of textual features exists in the study of electronic dominance, which is evident in social media and other electronic communications, to gain relative superiority (Kretinin, Samuel, & Kashyap, 2018). Text analysis and natural learning processing methods can be used to study dominance by looking at the use of upper-case letters, repeated letters, emoticons, and special characters. Kretinin, Samuel, and Kashyap (2018) found repetition of tweets by the same user, use of upper case, use of longer tweets, emphatic language, and counter posts upon analysis of tweets about certain stocks; all of these show dominant behavior of the user. Samuel, Holowczak, Benbunan-Fich, and Levine (2014) analyzed chat transcriptions to identify dominance using manual and automatic content analysis (Figure 2). They were able to show the potential for identifying dominance by using automatic textual analysis with generalized models for identifying dominance in computer-mediated group communication. Real time detection of dominance can provide feedback about group discussion nature (Samuel, Holowczak, Benbunan-Fich, & Levine, 2014).

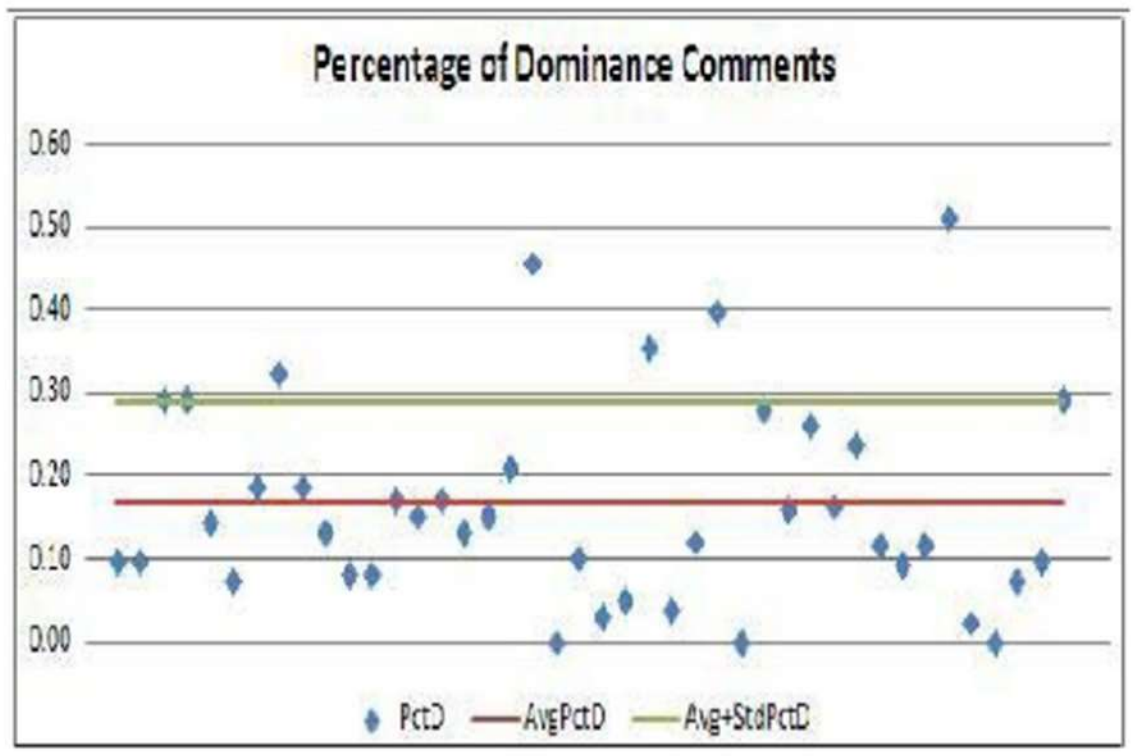

Figure 2: Dominance Comments Percentage Visualization (From Samuel, et., al., 2014, used with permission)

**Sense Visualization**

Sense Visualization (SViz) refers to the types of visual graphical representations of analysis that provide sense to textual characters, features or sequences. Common examples include: Sentiment analysis, Semantics and Natural Language Processing. A growing area of Natural Language Processing is sentiment analysis; there is research on this that spans from document analysis down to extremity of words and phrases (Kabir, Karim, Newaz & Hossain, 2018). Sentiment analysis is a method which organizes opinions to be classified into groups such as "Positive", "Negative", or sometimes "Neutral" and can also extract features from a text (Ashraf, Verma & Kavita, 2016). There are numerous approaches to sentiment analysis, as shown in studies on sentient analysis of Twitter data (Ashraf, Verma, and Kavita, 2016). Ortigosa, Martin, and Carro (2014) proposed a framework for sentiment analysis that starts when a user writes a message. This method looks to determine the sentiment polarity as a user posts and uses this to explore the user's emotions. Pak and Paroubek (2010) created a method to spontaneously obtain data that can be used for mining and sentiment analysis. They use the data to apply a sentiment classifier that can determine if the entire document is positive, negative, or neutral. Agarwal et al. (2011) considered two models for classification; the first of which considers only positive or negative classifications and the second of which also considers neutral classifications. They experimented with two types of models: a unigram model based on feature modeling and a tree 30 kernel-based model. From this they found that the tree kernel-based model outperforms other models. Aisopos, Papadakis, Tserpes, & Varvarigou, (2012) sought to resolve the lack of applicability of previous sentiment analysis by a polarity ratio and n-gram graphs for content-based information. They had found that these methods provided the sought-after effectiveness and were also language neutral and tolerant to noise; however, time efficiency should be improved for use with large data. Balahur et al. (2013) looked into the use of opinion mining to separate good and bad news from good and bad sentiment, as well as to look at dissimilar views on newspaper articles. Horakova (2015) used messages from social media sites to provide a view of business intelligence. Adarash and Kumar (2015) analyzed the occurrence of words and the use of emoticons to create sentiment and categorize users, they also used this sentiment to determine if a tweet was sent by a human or a bot. Sahayak, Shete, and Pathan (2015) used a machine learning algorithm that looked into the use of capitalization, internalization, emoticons, and negation handling neutralization to determine the sentiment of tweets. (Ashraf, Verma & Kavita, 2016).

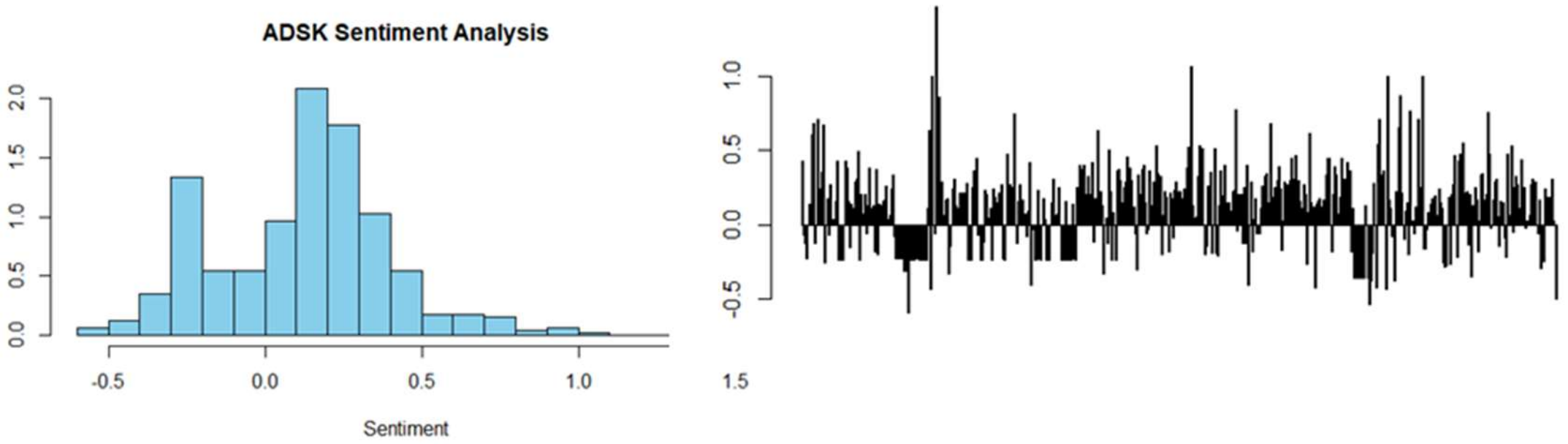

Figure 3: Autodesk Sentiment Analysis Visualization (From Kretinin, et., al., 2018, used with permission)

Kabir, Karim, Newaz, and Hossain (2018) looked to explore how feelings and sentiments are communicated in social media since the language is often casual and there can be restrictions on message length. They note that this area is not well studied and that the use of sentiment resources that were developed on non-microblogging information may not be as effective. They analyzed tweets from a famous Bangladeshi businessman and politician into positive, negative and unbiased sentiments by using R to create models. They were able to show that the user mostly posted neutral tweets and that the user posted positive tweets more than negative tweets. However, a limitation that they faced is that the twitter search API used to import the data can only obtain tweets from a maximum of seven days. They also note that sentiment analysis for twitter is not effective for detecting sarcasm and will detect it as negative sentiment; therefore computer categorization is not effective since the word "good" can mean good or bad depending on the context (Kabir, Karim, Newaz, & Hossain, 2018). Kretinin, Samuel, and Kashyap (2018) collected tweets associated with several stocks to evaluate the sentiment of the tweets. They classified tweets as positive, negative or neutral but did not include the neutral tweets in further analysis. They were able to show an alignment between sentiment trends and price movement for all four companies that were considered (Figure 3).

**Context Visualization**

Context Visualization (CViz) refers to the types of visual graphical representations of analysis that incorporate context to textual characters or sequences. They explore the "Who, When, and Where" of a social media posting as well as the modality being used. Users on social media are a part of two types of networks: a follower network and a reposting network (Chen, Lin, & Yuan, 2017). A user's follower network is based upon who they follow which is primarily based on relationships and interest while the reposting network is based upon what a user posts and reposts. Reposting of messages creates a diffusion of messages. A number of tools have been created to track a user's follower network. OntoVis is a tool that allows users to simplify networks and analyze relationships. A tool kit that can be used for network overview as well as discovery and exploration is NodeXL. MatrixExplorer can also be used to explore social media networks offering a node-link diagram as well as a matrix representation. One tool that is useful for finding common neighbors and large cliques is MatLink which offers a hybrid representation with links overlaid on a matrix. NodeTrix combines both MartixExplorer and MatLink to look at global and community structures. iO-LAP analyzes networks by looking at people, relation, content, and time. GraphDic allows for attributes of users such as age, gender, and location to easily compare. DemographicVis allows for an interactive analysis of features within demographic groups (Chen, Lin, & Yuan, 2017).

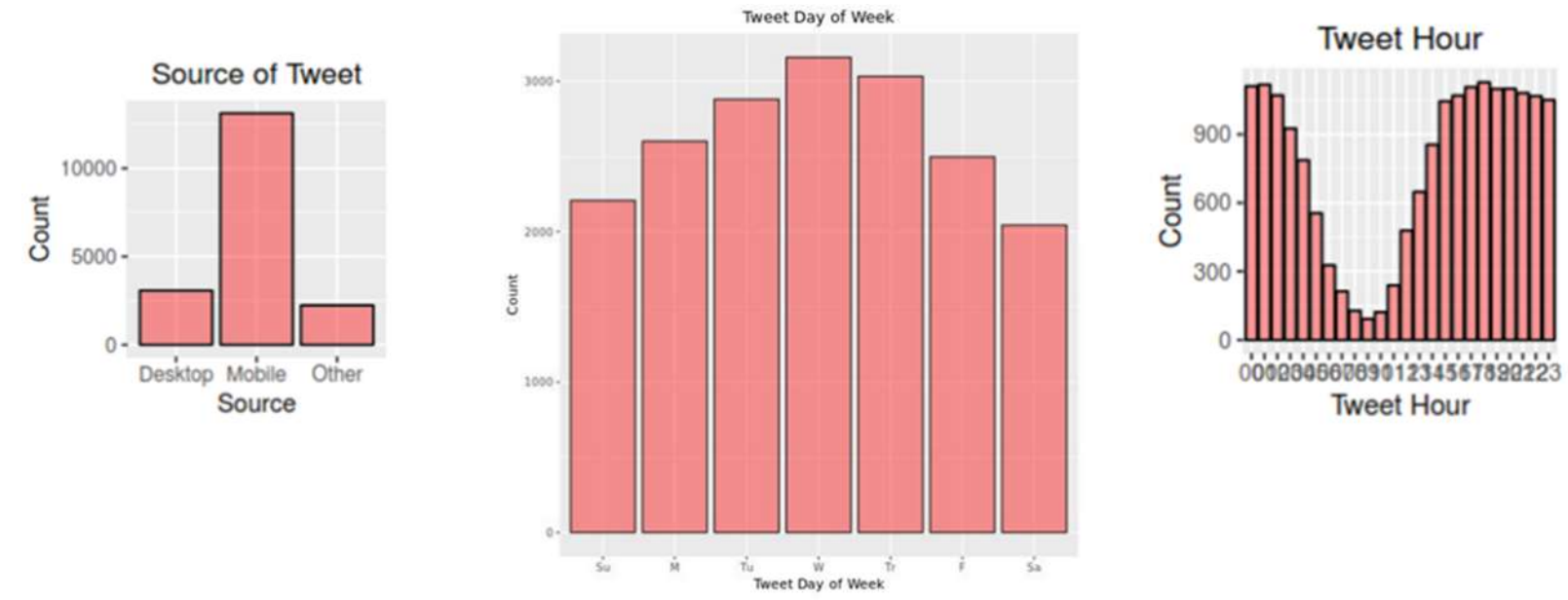

Figure 4: Viral Tweets Contextual Analysis Visualization (From Samuel, et., al., 2019, used with permission)

Reposting networks can also be visualized through a number of tools. One such tool is Google+Ripples which uses a combination of node-link and a circular map to show information flow (Chen, Lin, & Yuan, 2017). WeiboEvents provides the option of a tree layout, circular layouts, and a sail layout to explore the diffusion of information. One method to view anomalous information spread is FluxFlow. Users can also be defined through their geological location or their spatial temporal information. One of the earliest tools for this is Whisper, which uses a sunflower visualization to show how tweets spread from one source to the whole world. It is suggested that urban planning can benefit from the use of a circular visual design that uses land usage data and geo-tagged messages; while others use the geo-tagged data to find patterns between positions of interest (Chen, Lin, & Yuan, 2017). Twitter data lends itself to significant CViz as it is possible to associate over twenty contextual variables with tweets (Figure 4), creating a rich opportunity for gaining contextual insights associated with textual analytics of Tweets (Samuel, Garvey & Kashyap, 2019).

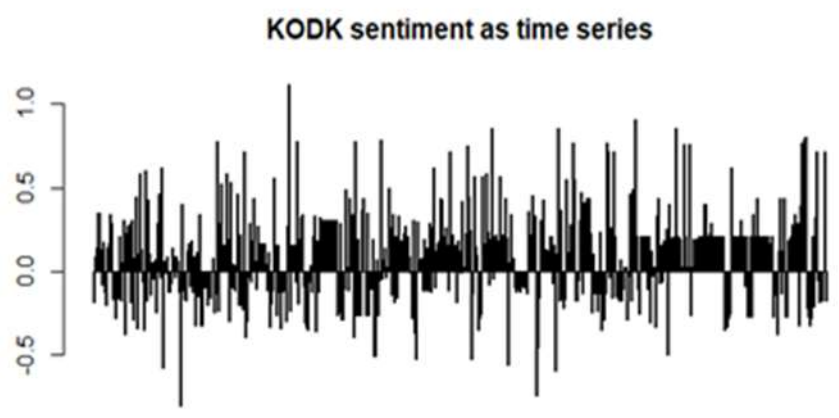

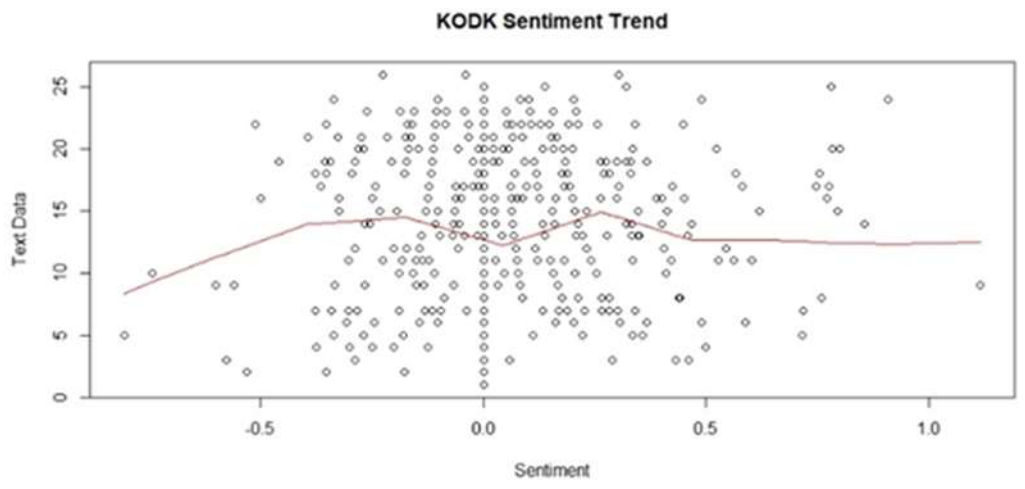

Figure 5 Figure 6

(Figure 5 & 6: Kodak Trend Analysis Visualization, From Kretinin, et., al., 2018, used with permission)

**Trend Visualization**

Trend Visualization (TViz) refers to the types of visual graphical representations of analysis that apply a timeline, process, or temporality to textual characters, features or sequences; this can be done through time series, process, or evolution. One tool to visualize the evolution of relationships is gestaltmatrix, which uses a glyph matrix for visualization (Chen, Lin, & Yuan, 2017). Textual data based sentiment analysis can also be visually represented as trends (Figure 5 & 6). A gestalt-based glyph can also be used to show relational data in chronological order. Graph flow can be used to show structural changes to a reposting network over time through the use of static flow visualization. Visual BackChannel uses a timeline to show keywords and topics; and a circular view to represent participants. ThemeCrowds creates a chronological layout of keywords through a hierarchical visualization. RoseRiver uses a tree-cut algorithm and adaptively finds suitable hierarchies for different times (Chen, Lin, & Yuan, 2017).

## 3 TEXTUAL DATA VISUALIZATION FRAMEWORK (TDViz)

Although there are data visualization frameworks, there is a need to develop a textual data visualization specific frame of reference because the properties, value addition and limitations of textual data are distinct. In an example of a typical big data domain relevant framework incorporating visualization, we reviewed a two-step framework for analyzing social media data that begins with preparing the data and finding an appropriate machine learning model to follow (Jimenez-Marquez, Gonzalez-Carrasco Lopez-Cuadrado, & Ruiz-Mezuca, 2019). The second step in their model uses big data architectures to find an outcome of the data using the machine learning model from the first step. Their framework is designed to process qualitative and quantitative information through Machine Learning and Natural Language Processing techniques. Their second stage comprises of several layers including a cluster resource manager layer, data access layer, data analysis layer, and a visualization layer. We propose a framework which is in contrast to such application-oriented frameworks, as a basis for classification of textual data visualization initiatives.

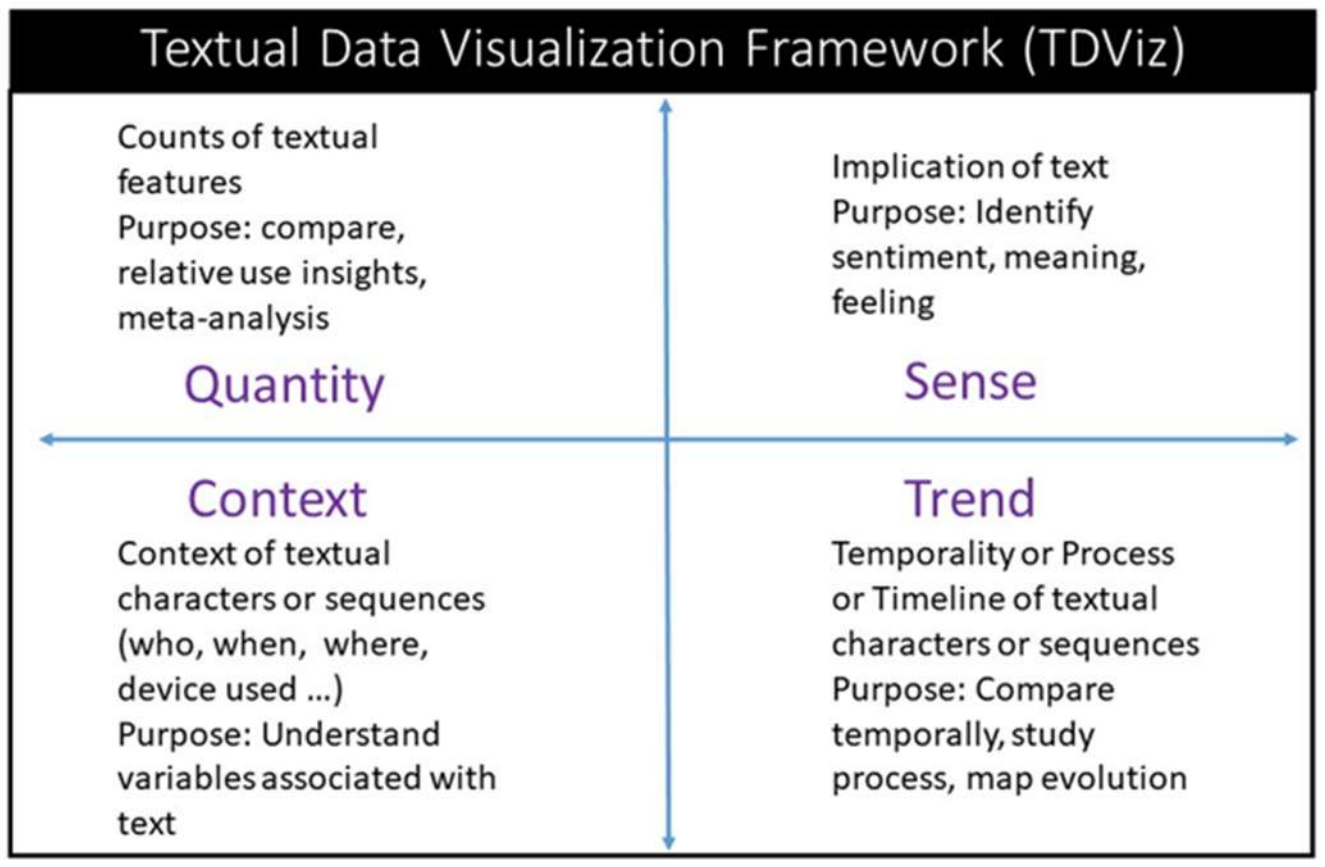

FIGURE 7: TDViz

## 4 DISCUSSION

This paper represents a part of our research on data visualization focused on textual analytics. This paper has some limitations – we have limited our study to static and electronically stored data. We have also avoided discussion on the richness of data and the impact of information facets as it has been shown that the characteristics of information can influence performance (Samuel, 2017) and therefore potentially the sentiment associated with textual information. However, these boundaries also open up opportunities for further research into the excluded knowledge areas.

Much of current data visualization is increasingly tool driven and therefore it is important to understand the data visualization tools environment. Gartner has identified Tableau to be an industry leader in data visualization for the past six years from 2013 to 2018, and it is expected to retain its position for 2019 (Tableau website, 2019). Tableau lends itself to easy and efficient creation of data visualizations. However, in our limited experience, we find R and its data visualization packages (libraries) to provide more options, variety and customizability, and therefore to be the best data visualization tool for serious work requiring greater control of visualizations and flexibility to create dynamic plots and special visualizations such as three-dimensional representations with superior customizability (R-project.org, 2019). Other tools, including Splunk have been recommended for students to learn to visualize data. Splunk is a business intelligence tool that can be used for collecting storing and analyzing data. Its main advantages are that is has a free version, it has an intuitive graphical interface, it can be used for analysis of streaming data, and it can be used with almost any type of data. Sigman et al. (2016) used REST API to bring twitter data into Splunk, allowing the user to create functions to determine what data they want to bring in and allowing for data filtering based on keywords and hashtags. Splunk also has a driver that allows Tableau to use data directly from Splunk. Research has identified the benefit of using parsimonious tools like Tableau and Splunk to be student friendly, allowing exploration of big data without having to worry about technical issues and nuances (Sigman, et al., 2016).